%% file: main.tex
\documentclass[conference]{IEEEtran}
\IEEEoverridecommandlockouts
\usepackage{cite}
\usepackage{amsmath,amssymb,amsfonts}
\usepackage{graphicx}
\usepackage{mathtools}
\usepackage{amsmath}
\usepackage{textcomp}
\usepackage{xcolor}
\usepackage{subcaption}
\usepackage{multirow} 
\usepackage{tablefootnote}
\usepackage[numbers,sort&compress]{natbib}
\def\BibTeX{{\rm B\kern-.05em{\sc i\kern-.025em b}\kern-.08em
    T\kern-.1667em\lower.7ex\hbox{E}\kern-.125emX}}

\makeatletter 
\newcommand{\linebreakand}{%
  \end{@IEEEauthorhalign}
  \hfill\mbox{}\par
  \mbox{}\hfill\begin{@IEEEauthorhalign}
}
\makeatother 
\begin{document}

\title{Benchmarking Emerging Cavity-Mediated Quantum Interconnect Technologies for Modular Quantum Computers
\thanks{We extend our sincere gratitude to Prof. Peter Haring- Bol\'ivar for his valuable insights concerning the proposed figure of merit, which significantly contributed to this work. We gratefully acknowledge funding from the European Commission through HORIZON-EIC-2022-PATHFINDEROPEN-01-101099697 (QUADRATURE) and grant HORIZON-ERC-2021-101042080 (WINC). Authors acknowledge support from the QCOMM-CAT-Planes Complementarios: Comunicacion Cuántica - supported by MICIN with funding from the European Union,  NextGenerationEU (PRTR-C17.I1) and by Generalitat de Catalunya, and by ICREA Academia Award 2024. CGA also acknowledges funding from the Spanish Ministry of Science, Innovation and Universities through the Beatriz Galindo program 2020 (BG20-00023).}
}

\author{
    \IEEEauthorblockN{Sahar Ben Rached}
    \IEEEauthorblockA{\textit{NanoNetworking Center in Catalunya} \\
    \textit{Universitat Polit\`ecnica de Catalunya}\\
    Barcelona, Spain \\
    sahar.benrached@upc.edu}
    \and
    \IEEEauthorblockN{Sergio Navarro Reyes}
    \IEEEauthorblockA{\textit{NanoNetworking Center in Catalunya} \\
    \textit{Universitat Polit\`ecnica de Catalunya}\\
    Barcelona, Spain \\
    sergio.navarro.reyes@estudiantat.upc.edu}
    \and
    \IEEEauthorblockN{Junaid Khan}
    \IEEEauthorblockA{\textit{NanoNetworking Center in Catalunya} \\
    \textit{Universitat Polit\`ecnica de Catalunya}\\
    Barcelona, Spain \\
    junaid.khan@upc.edu}
    \and
    \IEEEauthorblockN{Carmen G. Almud\'ever}
    \IEEEauthorblockA{\textit{Computer Engineering Department} \\
    \textit{Universitat Polit\`ecnica de Val\`encia}\\
    Val\`encia, Spain \\
    cargara2@disca.upv.es}
    \and
    \IEEEauthorblockN{Eduard Alarc\'on}
    \IEEEauthorblockA{\textit{NanoNetworking Center in Catalunya} \\
    \textit{Universitat Polit\`ecnica de Catalunya}\\
    Barcelona, Spain \\
    eduard.alarcon@upc.edu}
    \and
    \IEEEauthorblockN{Sergi Abadal}
    \IEEEauthorblockA{\textit{NanoNetworking Center in Catalunya} \\
    \textit{Universitat Polit\`ecnica de Catalunya}\\
    Barcelona, Spain \\
    abadal@ac.upc.edu}
}

\maketitle

\input{abstract}
\input{introduction}
\input{methodology}

\input{results}

\input{discussion}
\input{conclusion}

\input{bibliography}
\end{document}

%% file: abstract.tex
\begin{abstract}
Modularity is a promising approach for scaling up quantum computers and therefore integrating higher qubit counts. The essence of such architectures lies in their reliance on high-fidelity and fast quantum state transfers enabled by generating entanglement between chips. In addressing the challenge of implementing quantum coherent communication channels to interconnect quantum processors, various techniques have been proposed to account for qubit technology specifications and the implemented communication protocol. By employing Design Space Exploration (DSE) methodologies, this work presents a comparative analysis of the cavity-mediated interconnect technologies according to a defined figure of merit, and we identify the configurations related to the cavity and atomic decay rates as well as the qubit-cavity coupling strength that meet the efficiency thresholds. We therefore contribute to benchmarking contemporary cavity-mediated quantum interconnects and guide the development of reliable and scalable chip-to-chip links for modular quantum computers.
\end{abstract}

%% file: introduction.tex
\section{Introduction}
\begin{table*}
    \centering
    \begin{tabular}{ccccc}
    \hline
        \textbf{Qubit type} & \textbf{Reference} & \textbf{Coupling strength \( g \) (Hz)} & \textbf{Cavity decay rate \( \kappa \) (Hz)} & \textbf{Atomic decay rate \( \gamma \) (Hz)} \\
    \hline
        \textbf{Superconducting} & \textbf{Magnard et al. \cite{b16}} & \(307 \cdot 10^6\) & \(8.6 \cdot 10^6\) &  \\
    \hline
        \multirow{3}{*}{\textbf{Neutral Atom}} & \textbf{Ramette et al. \cite{b17}} & \(5.8 \cdot 10^6\) & \(0.34 \cdot 10^6\) & \(6 \cdot 10^6\) \\
        & \textbf{Young et al. \cite{b18}} & \(98 \cdot 10^6\) & \(253 \cdot 10^6\) & \(6 \cdot 10^6\) \\
        & \textbf{Liu et al. \cite{b19}} & \(3.2 \cdot 10^6\) & \(10^6\) & \(2.6 \cdot 10^6\) \\
    \hline
       \textbf{Trapped ion} & \textbf{Ramette et al. \cite{b17}} & \(2.8 \cdot 10^6\) & \(5.3 \cdot 10^4\) & \(25 \cdot 10^6\) \\
    \hline
        \multirow{4}{*}{\textbf{Semiconducting}} & \textbf{Sipahigil and Evans et al. \cite{b20}} & \(2.1 \cdot 10^9\) & \(57 \cdot 10^9\) & \(0.3 \cdot 10^9\) \\
        & \textbf{Bonizzoni et al. \cite{b21}} & \(21 \cdot 10^6\) & \(10 \cdot 10^6\) & \(30 \cdot 10^6\) \\
        & \textbf{Schuster et al. \cite{b22}} & \(38 \cdot 10^6\) & \(1.3 \cdot 10^6\) & \(96 \cdot 10^6\) \\
        & \textbf{Evans et al. \cite{b23}} & \(7.3 \cdot 10^9\) & \(48 \cdot 10^9\) & \(0.19 \cdot 10^9\) \\
    \hline
    \end{tabular}
    \caption{Selected references for the comparative analysis among contemporary interconnect technologies. The atomic (qubit) decay rate $\gamma$ in \cite{b16} is not listed.}
    \label{references}
\end{table*}

Modular quantum computing architectures are proposed as a viable approach to scale current quantum computing systems to host a large number of qubits to solve practical problems. This approach is based on interconnecting smaller-sized quantum processors (also referred to as \emph{cores}) to potentially alleviate critical technical challenges associated with packing a high number of qubits within a single chip. The main challenges addressed are the effects of crosstalk, quantum state disturbance, and the increased complexity of the control system \cite{b1} that significantly degrade the computational performance.

Building reliable and scalable modular quantum computing systems is contingent on integrating fast, high-fidelity inter-core operations to perform quantum state transfer and entanglement generation across cores. In recent years, several experimental works aiming to integrate quantum coherent networks have emerged for the various qubit technologies, such as the invention of SNAIL-based quantum state routers \cite{b2}, integrated photonic networks \cite{b3}, and modeling the integration of entanglement distillation for EPR pair-mediated protocols \cite{b4}, in addition to advancing techniques to enhance the scalability and robustness of on-chip networks such as the loss suppression techniques to extend qubits' coherence times \cite{b5} and transduction schemes \cite{b6}. The efficiency of quantum interconnects is measured by various experimental parameters, including cooperativity, success probability, infidelity, and operation duration, assessing the performance of the quantum communication process. Nevertheless, current proposals do not meet the efficiency threshold that guarantees their incorporation within large-scale systems, as typically, such systems are characterized by a low process fidelity \cite{b7}, prohibitively long latencies \cite{b8}, or both \cite{b9}.

Chip-to-chip interconnect techniques depend on the qubit technology and the implemented protocol. This has led to the emergence of a variety of methods, each tailored to optimize the quantum state transfer and entanglement generation across cores. Teleportation-based techniques are prevalent approaches aiming to operate non-local gates necessary for the distribution of quantum tasks across cores by utilizing entanglement to connect modules \cite{b10}. The primary challenge of these systems is enhancing the entanglement generation rate and the inter-core operations to be comparable with the rate of local entangling gates and operations within each module. This adjustment is critical for minimizing the overhead associated with distributing quantum tasks, essentially in terms of operation fidelity and latency, therefore improving the overall system efficiency. 

Alternatively, optical cavities and waveguides are employed to transport quantum states between qubits placed in different cores \cite{b11}. Ensuring the reliability of these interconnects accounts for the qubit technology and the transmission line; where the focus is on strengthening light-matter interactions to enable effective quantum state transfers. For optimal performance, these structures must support high-quality factors and small mode volumes to maximize the interaction strength, and are typically designed to operate at short wavelengths and maintain a low-loss process at cryogenic temperatures. Integrated photonic networks, which are particularly useful for trapped ion platforms \cite{b12}, provide efficient routing of photons across cores while maintaining a high-fidelity multiphoton interface. 

Cavities are strong candidates to build quantum networks given their versatility and advantageous feature of preserving photons for longer times, enabling their coherent transmission with limited dissipation. Hence, within the current developmental stage of inter-core communication technologies, we apply Design Space Exploration (DSE) methodologies to benchmark cavity-mediated interconnect technologies and define optimal regions of performance of critical parameters of the system, including the atomic (qubit) decay rate $\gamma$, cavity decay rate $\kappa$, as well as the qubit-cavity coupling strength $g$, according to a proposed figure of merit that encapsulates the success probability of the quantum state transfer, process latency, and infidelity. Our contribution consists of a comparative analysis that evaluates contemporary proposals of quantum interconnect technologies and forecasts potential enhancements needed to meet and surpass the efficiency thresholds.

\begin{figure*}
    \centering
    \begin{subfigure}{.32\textwidth}
      \centering
      \includegraphics[width=\linewidth]{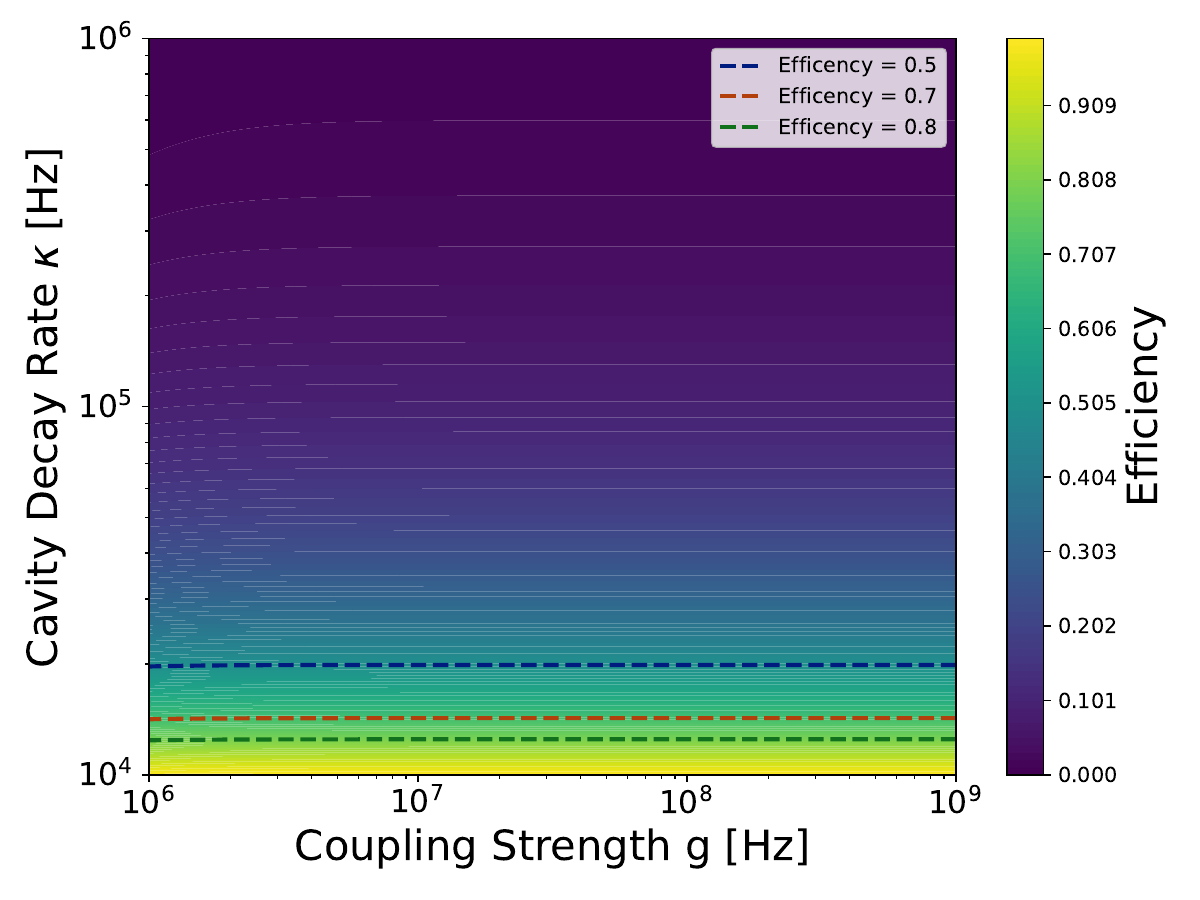}
      \caption{Delineating the design requirements for $g$ and $\kappa$ achieving state transfer success probabilities of $0.5$, $0.7$, and $0.8$.}
      \label{efficiency}
    \end{subfigure}%
    \hspace{0.1cm}
    \begin{subfigure}{.32\textwidth}
      \centering
      \includegraphics[width=\linewidth]{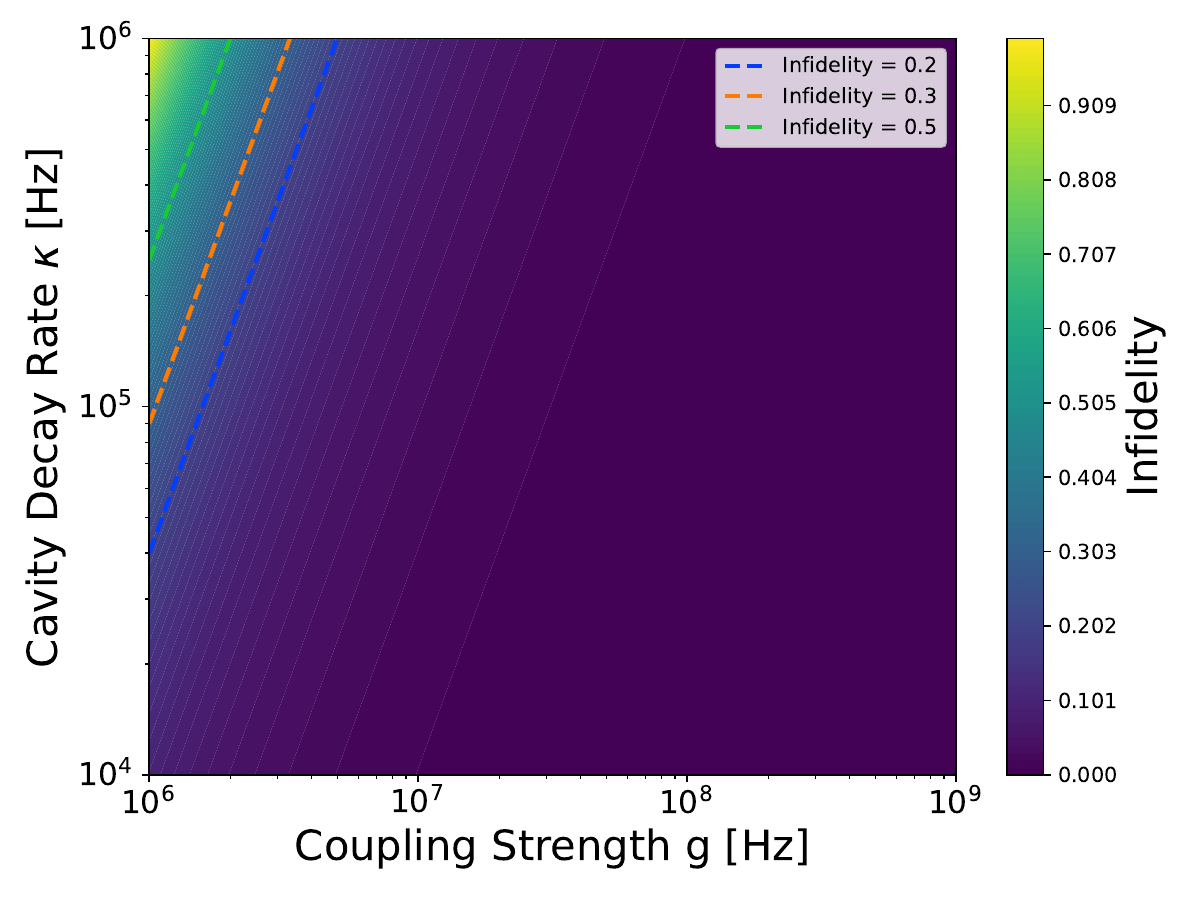}
      \caption{Delineating the design requirements for $g$ and $\kappa$ to achieve infidelity rates of $0.2$, $0.3$, and $0.5$.}
      \label{infidelity}
    \end{subfigure}%
    \hspace{0.1cm}
    \begin{subfigure}{.32\textwidth}
      \centering
      \includegraphics[width=\linewidth]{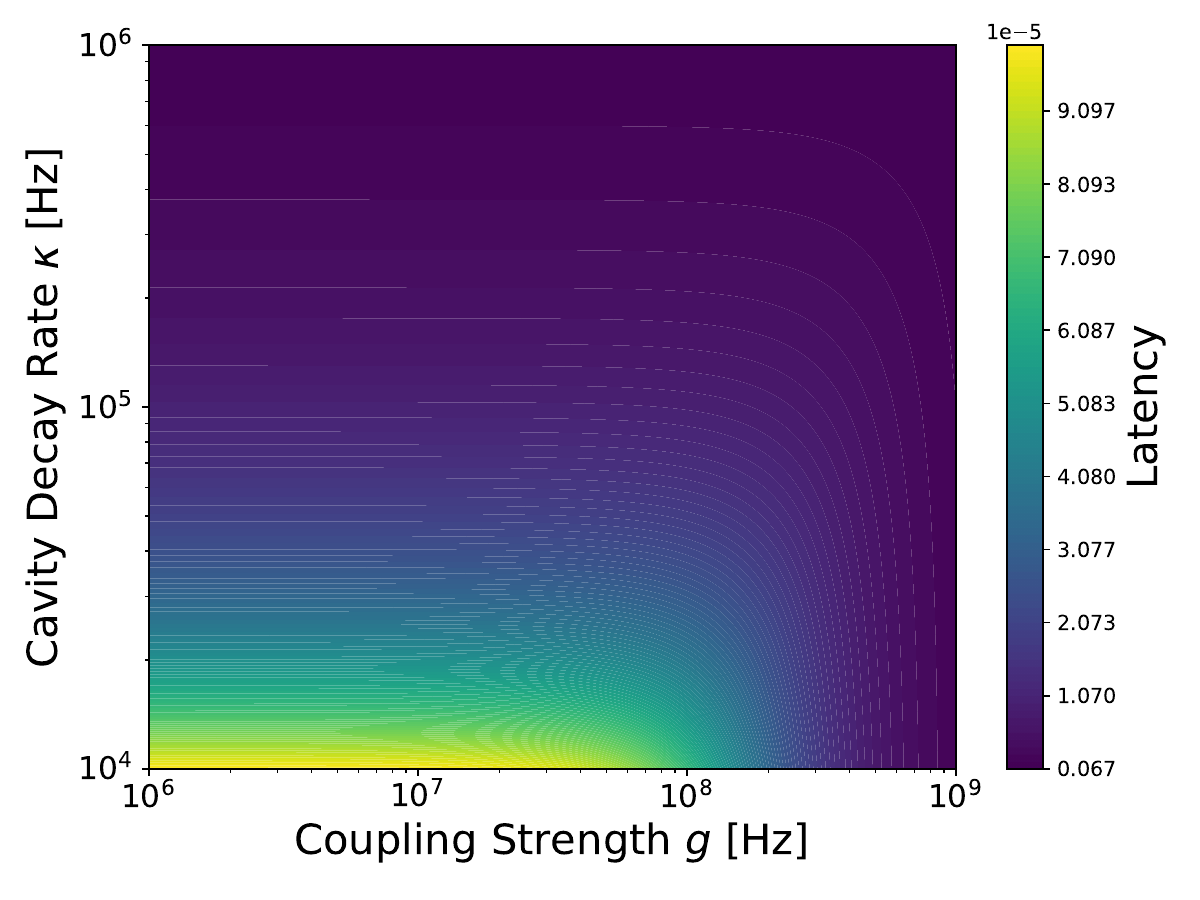}
      \caption{Delineating the design requirements for $g$ and $\kappa$ for minimized operation latency.\\}
      \label{latency}
    \end{subfigure}%
    \caption{Analysis of the design space exploring the variation of the (a) efficiency, (b) infidelity, and (c) latency to the coupling strength $g$ and the cavity decay rate $\kappa$. We consider $\kappa_{ex} = 10^6$ Hz, $\gamma = 10^6$ Hz, and $I_g' = 1$.}
    \label{fig:design_space_analysis}
\end{figure*}

%% file: methodology.tex
\section{Methodology}
We apply DSE methodologies to assess quantum interconnects and extrapolate on the optimal design requirements based on our proposed a figure of merit $\Gamma$ in Eq. \ref{figure} that integrates multiple performance metrics evaluating the efficacy of inter-core operations between distant qubits (i.e placed in different cores). This figure of merit encapsulates three crucial aspects: the efficiency of quantum state transfer modeled by the success probability, the operation latency, and the infidelity of the process.

We define the figure of merit $\Gamma$ as follows:
    \begin{equation}\label{figure}
        \Gamma =  \frac{\alpha \times \text{Efficiency}}{(1-\alpha) \times (\text{Latency} \times \text{Infidelity})}
    \end{equation} 
where $\alpha$ is a weighting factor that adjusts the relative importance of efficiency versus the product of latency and infidelity.

The efficiency of the quantum state transfer is modeled as the success probability $P_s$, which is a suggested figure of merit in \cite{b13} applicable to deterministic single-photon generation schemes in cavity-QED systems:
    \begin{equation}
        \text{Efficiency} \propto P_s = (\frac{\kappa_{ex}}{\kappa}) (\frac{2C}{1+2C}) (1 - \frac{I_g'}{\kappa C})
    \end{equation}
where $\kappa_{ex}$ represents the external cavity decay rate, $\kappa$ is the decay rate of the cavity, $C$ denotes the cooperativity parameter, and $I_g'$ quantifies inherent imperfections or losses in the system.

\begin{figure*}
    \centering
    \begin{subfigure}{.43\textwidth}
      \centering
      \includegraphics[width=1\textwidth]{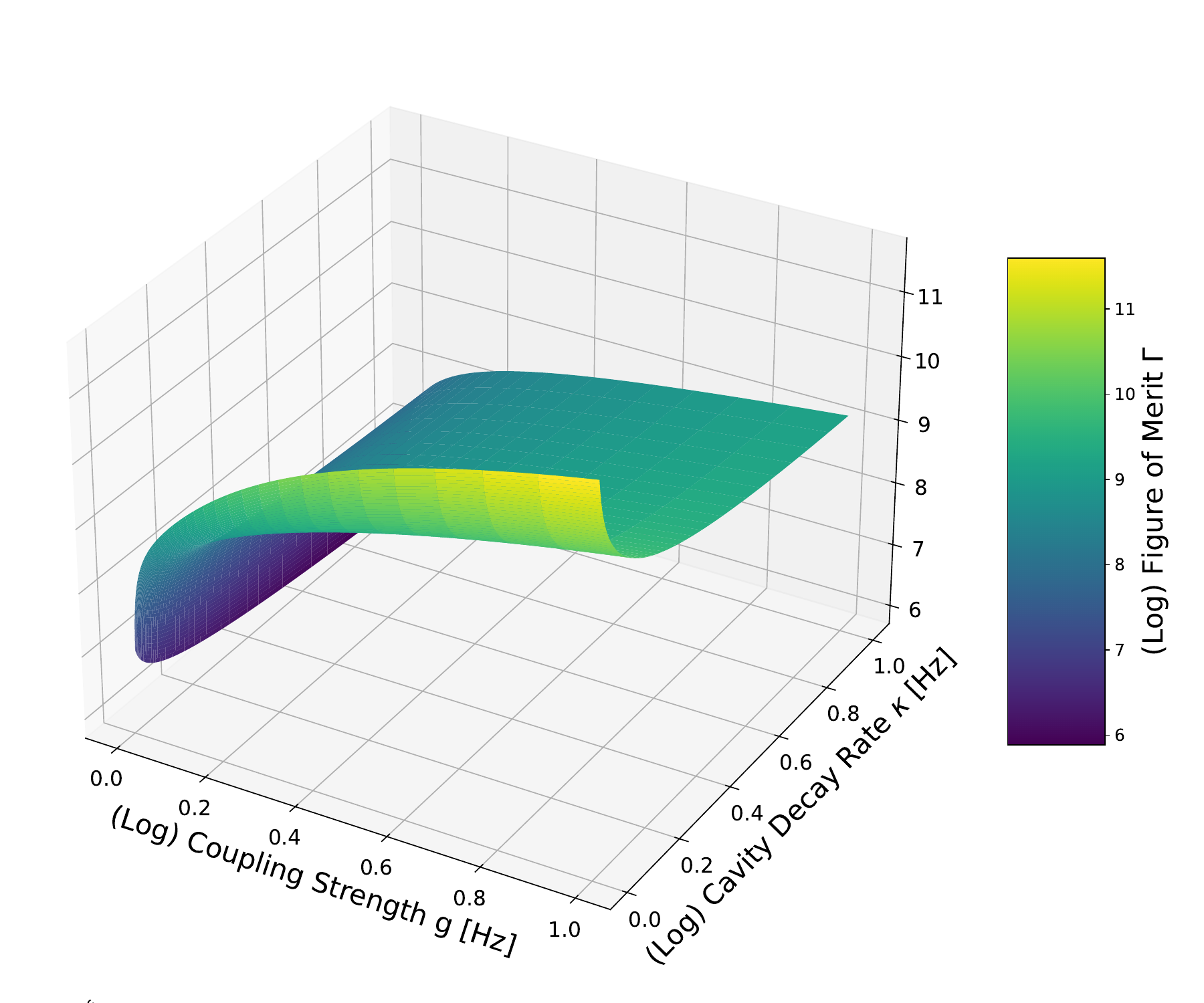}
      \caption{}
      \label{FM_gk_3D}
    \end{subfigure}%
    \begin{subfigure}{.5\textwidth}
      \centering
      \includegraphics[width=1\textwidth]{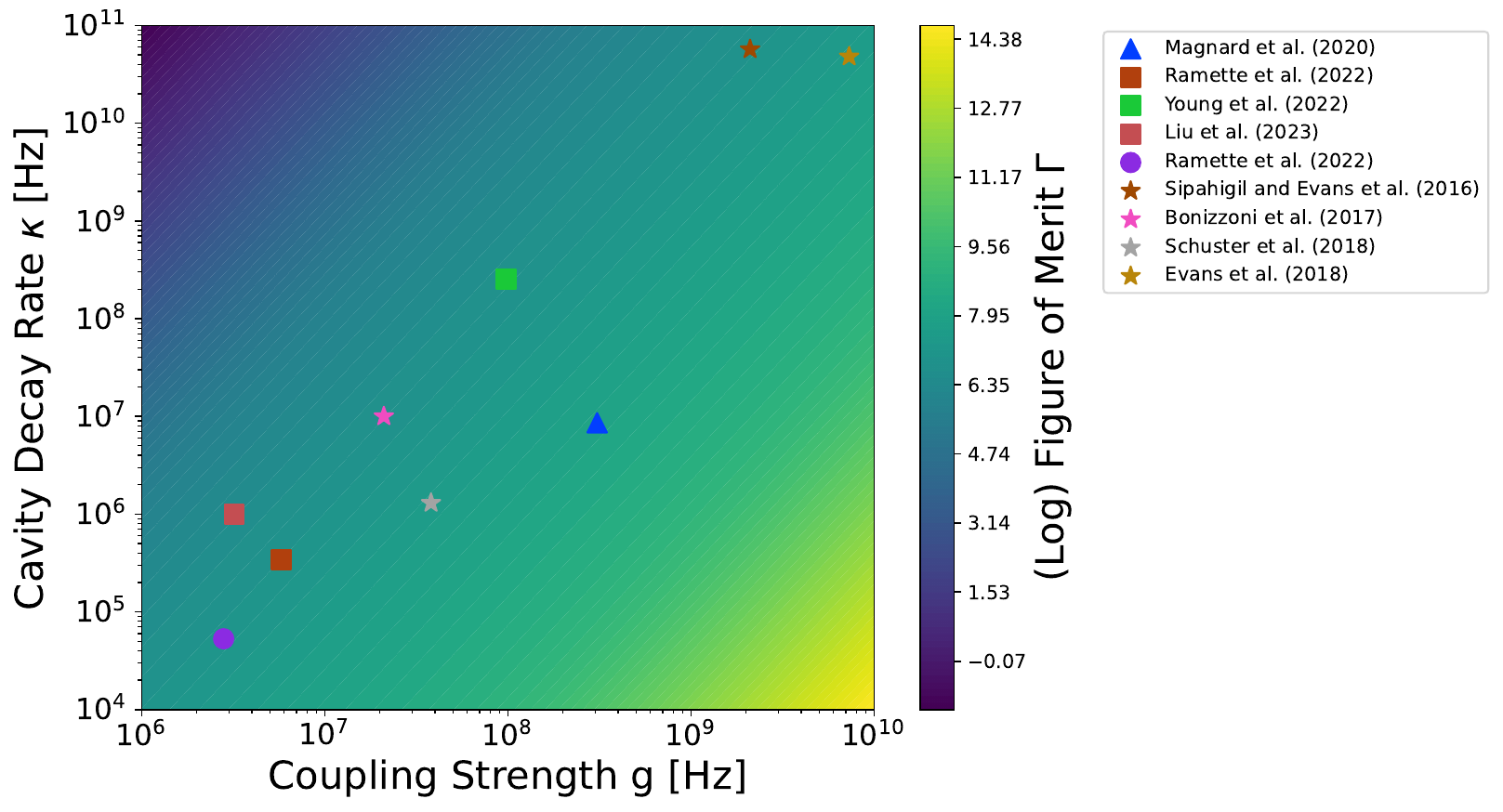}
      \caption{}
      \label{FM_gk_2D}
    \end{subfigure}%
    \caption{The variation of the figure of merit $\Gamma$ with respect to the coupling-strength $g$ and the cavity decay rate $\kappa$ (a) within the design space (b) showcasing a comparative analysis of selected state-of-the-art quantum interconnect technologies designed for superconducting (triangle) \cite{b16}, semiconducting (star) \cite{b17,b18,b19}, neutral atom (square) \cite{b20,b21,b22,b23}, and trapped ion (circle) \cite{b17} qubit technologies. We consider the remaining parameters $\kappa_{ex} = 10^6$ Hz, $\gamma = 10^6$ Hz, $I_g' = 1$, and $\Delta = 2\cdot10^9$  Hz.}
\label{variation_of_FM_gk}    
\end{figure*} 

The cooperativity parameter $C$ is defined as:
    \begin{equation}
        C = \frac{g^2}{\kappa \gamma}
    \end{equation}
with $g$ being the coupling strength between the qubit and the cavity, and $\gamma$ is the inherent loss rate of the qubits. The cooperativity $C$ is employed to evaluate the qubit-cavity coupling regime, where $C > 1$ indicates the coupling is strong, meaning that at resonance, nearly every photon entering the cavity is coherently transferred into the target qubit.

The infidelity ($1 - F$) inversely reflects the fidelity of the quantum state transfer, modeled as:
    \begin{equation}
        \text{Infidelity} \propto \sqrt\frac{\kappa \gamma}{g^2}
    \end{equation}
as proposed in \cite{b14} for atomic qubits connected to cavities.

In the context of quantum interconnects, the latency associated with the transfer of a quantum state is an essential factor to evaluate the performance of the system. Latency in such systems can be represented by the ratio of the superconducting cavity quality factor $Q$ to the cavity frequency $\omega_c$. Here, the quality factor relates to the number of oscillation cycles required for the energy in the cavity to decay. Hence, the defined ratio characteristically represents the number of oscillation cycles by the duration of a single oscillation cycle, and therefore the ratio $\frac{Q}{\omega_c}$ \cite{b15} represents the time necessary to transfer a quantum state through the cavity:
    \begin{equation}
        \text{Latency} \propto \frac{Q}{\omega_c} = \frac{\Delta^2 + \gamma^2}{2 g_{eff,s}^2 \gamma + \kappa(\Delta^2 + \gamma^2)}
    \end{equation}
where $\Delta$ represents the detuning between the qubit and the resonant frequency of the cavity. We denote that a high quality factor translates to higher latency.

For the comparative analysis of state-of-the-art interconnect technologies presented throughout this paper, we have selected nine references to experimental works using cavity-QED systems to perform quantum state transfers between distant qubits as listed in Table \ref{references}. The selected works provide experimental values for the fundamental parameters governing our proposed figure of merit: the coupling strength $g$, cavity decay rate $\kappa$, and atomic decay rate $\gamma$.

%% file: results.tex
\section{Results}
\begin{figure*}
    \centering
    \begin{subfigure}{.43\textwidth}
      \centering
      \includegraphics[width=1\textwidth]{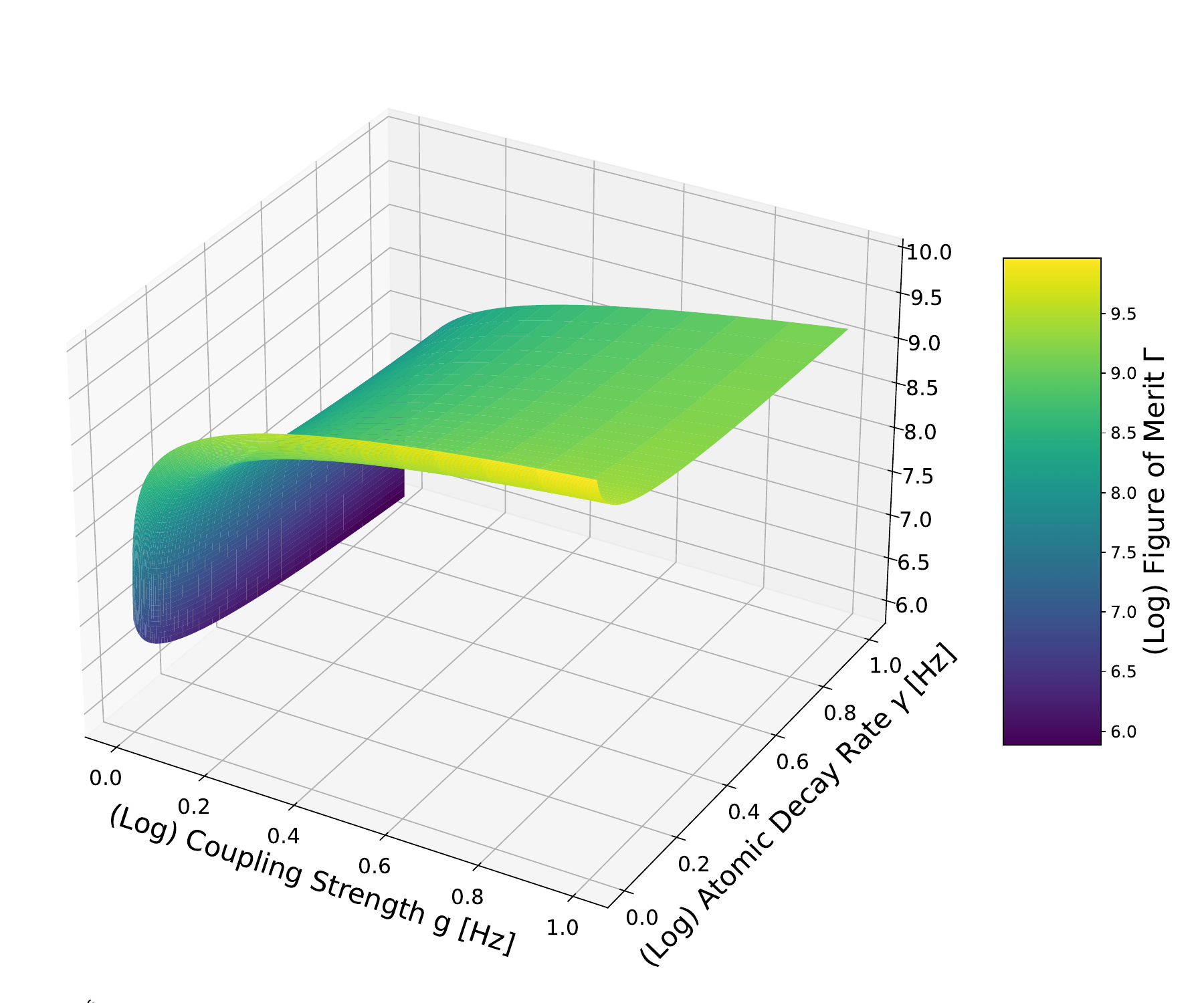}
      \caption{}
      \label{FM_ga_3D}
    \end{subfigure}%
    \begin{subfigure}{.5\textwidth}
      \centering
      \includegraphics[width=1\textwidth]{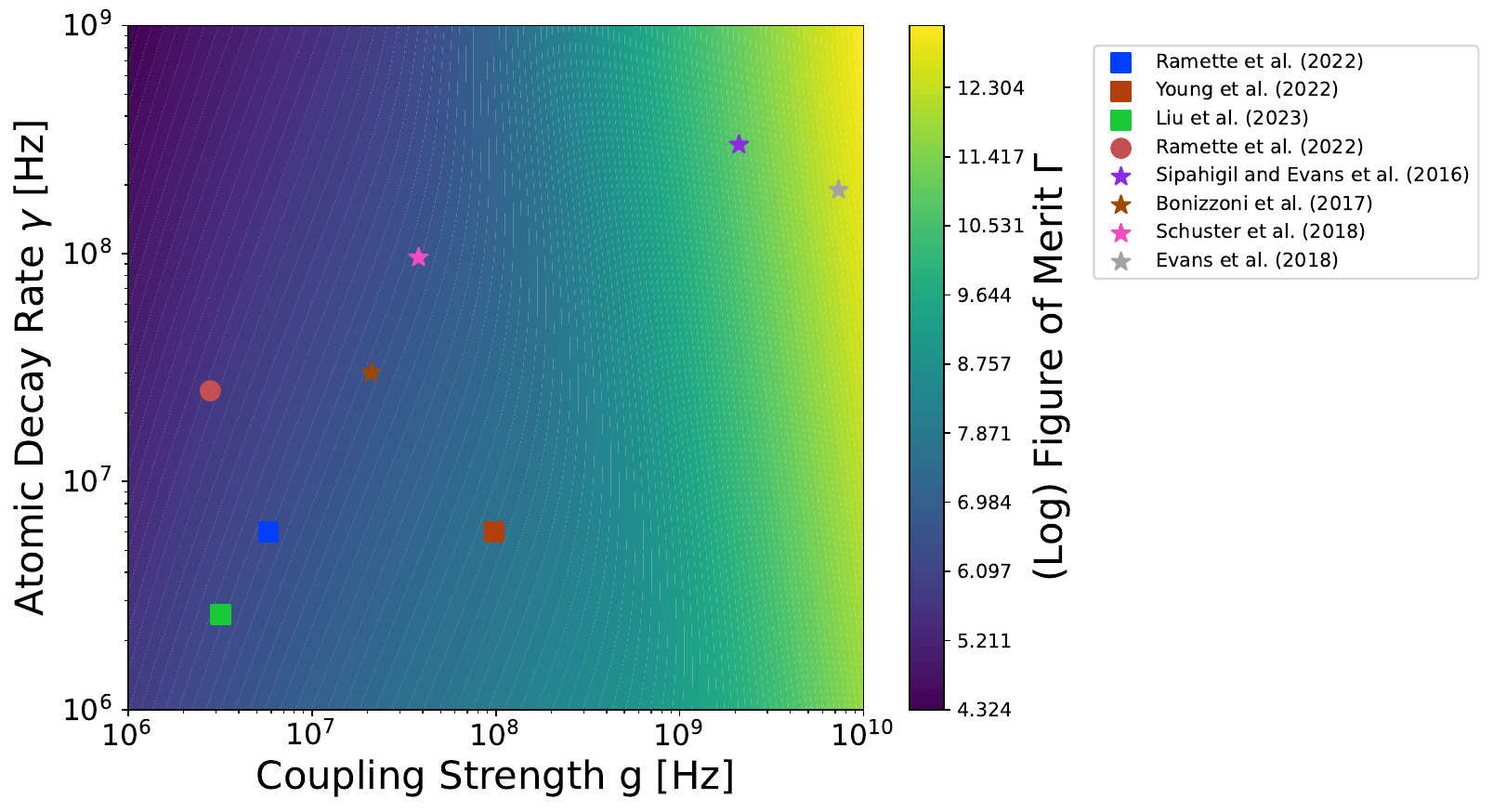}
      \caption{}
      \label{FM_ga_2D}
    \end{subfigure}%
    \caption{The variation of the figure of merit $\Gamma$ with respect to the coupling-strength $g$ and the atomic decay rate $\gamma$ (a) within the design space (b) showcasing a comparative analysis of selected state-of-the-art quantum interconnect technologies designed for semiconducting (star) \cite{b17,b18,b19}, neutral atom (square) \cite{b20,b21,b22,b23}, and trapped ion (circle) \cite{b17} qubit technologies. We consider the remaining parameters $\kappa_{ex} = 10^6$ Hz, $\kappa = 10^6$ Hz, $I_g' = 1$, and $\Delta = 2.10^9$ Hz.}
\label{variation_of_FM_ga}
\end{figure*}

The analysis of the decay mechanisms and the interaction strength in the qubit-cavity systems is important to understand the decoherence process, which significantly impacts the efficiency of quantum computations. This analysis examines the trade-offs of the qubit-cavity coupling strength $g$, and decay rates $\kappa$ and $\gamma$ based on the proposed figure of merit with the objective of identifying optimal performance regions that maximize efficiency while concurrently minimizing latency and infidelity.

To conduct the DSE based on the proposed figure of merit, if one would want to prioritize either maximizing efficiency or minimizing the effects of transfer latency and infidelity, this can be adjusted with the weighting parameter $\alpha$. Our work focuses on delineating the optimal performance regions for cavity-mediated interconnects considering a trade-off between the interconnect efficiency, and the impact of process latency and infidelity. Hence, in the following DSE and comparative analyses, we consider $\alpha = 0.5$.

In Figure \ref{efficiency}, we showcase the dependence of the efficiency on $g$ and $\kappa$. We note that within the strong coupling regime ($g \gg (\kappa, \gamma)$), where $g$ significantly exceeds $\kappa$ and $\gamma$, we observe a peak in efficiency when the cavity decay rate is minimized and confined within a narrow range of $10^4$ Hz across all the considered coupling strength ranges. In Figure \ref{infidelity}, infidelity shows minimal values at low $\kappa$ ranges within the strong coupling regime. However, as the coupling strength diminishes to intermediate ($\kappa  \thickapprox g^2 / \kappa \gg \gamma$) or weak regimes ($\kappa  \gg g^2 / \kappa \gg \gamma$), infidelity surges to reach a rate of $0.9$. The latency as analyzed in Figure \ref{latency} indicates that the inter-core operation time is inversely related to $g$, as higher rates of $g$ substantially reduce latency across various $\kappa$ rates. This correlation shows the importance of maintaining high coupling strengths to counteract high rates of $\kappa$ and ensuring rapid quantum state transfers. These results highlight the specific, narrow ranges of $\kappa$ and $g$ where the efficiency, infidelity, and latency reach their optimal values, pointing to the need for accurate designs of quantum interconnects.

In Figures \ref{variation_of_FM_gk} and \ref{variation_of_FM_ga}, we present a comparative analysis of state-of-the-art quantum interconnect technologies developed for superconducting, semiconducting, trapped ions, and neutral atom qubit technologies. Using the defined figure of merit, we examine its dependence on the coupling strength $g$ and the cavity decay rate $\kappa$, shown in Figure \ref{variation_of_FM_gk}. This analysis suggests that the optimal performance of cavity-mediated interconnects is achieved when the coupling strength $g$ exceeds $10^8$ Hz, and the cavity decay rate $\kappa$ remains below $10^6$ Hz, corresponding to a strong coupling regime that effectively mitigates the adverse effects of high cavity decay rates to sustain the system's performance. In Figure \ref{FM_gk_2D}, we compare the performance of the nine selected early-stage experimental works in terms of system coupling strength and the induced cavity loss rates. Although the evaluated proposals fall within the intermediate to near-strong coupling regimes —currently insufficient for scalable and reliable networks as indicated by the previous results— we note a significant progress over the years in reducing cavity decay rates compared to initial proposals. 

In Figure \ref{variation_of_FM_ga}, we showcase the dependence of the proposed figure of merit on the qubit decay rate $\gamma$ and qubit-cavity coupling strength $g$. The importance of enhancing the coupling strength is further emphasized for the purpose of improving the system's overall performance. The comparative analysis depicted in Figure \ref{FM_ga_2D} demonstrates that while the reviewed interconnect technologies are primarily classified within the intermediate to near-strong coupling regimes, there has been a significant reduction in qubit decay rates over the recent years, and the technical challenges remain to minimize qubit losses and concurrently enhance coupling to the cavity. 

The presented comparative analysis reveals that state-of-the-art proposals are characterized by either prohibitively high values of $\kappa$ and $\gamma$, or insufficiently low values of $g$. This positioning indicates that despite significant technological advancements, current quantum interconnect proposals still have a substantial room for improvement to enhance efficiency and expand inter-core communication networks effectively. 

Considering that the coupling strength $g$ is a deterministic factor to maximize the figure of merit, we perform a sensitivity analysis in Figure \ref{sa1} to understand the robustness of quantum interconnects against intrinsic decoherence mechanisms. We highlight the threshold of the strong coupling regime in each case, below which the interconnect performance is suboptimal. For lower $\kappa$ - $\gamma$ ranges, the figure of merit increases more steeply, indicating that interconnect technologies operating within the weak and intermediate coupling regimes are highly sensitive to variations in the coupling strength and are, therefore, more capable of capitalizing on the enhancements in $g$ for performance improvements. Higher $\kappa$ - $\gamma$ ranges exhibit a less pronounced sensitivity, which indicates approaching a saturation stage, which may be attributed to the degrading effects of inherent losses in the system at higher decay rates.

This analysis, besides providing a benchmarking framework for emerging quantum interconnects, potentially contributes to advancing architectures of modular quantum computers by providing an overall system view accounting for the specifications of the qubits, interconnects, and the dynamics of their interactions.

%% file: discussion.tex
\section{Discussion}
In this work, we present a comprehensive analysis of decay rates and the coupling strength in cavity-mediated quantum interconnects for inter-core quantum state transfer in modular quantum computing systems. We aim to benchmark state-of-the-art interconnect techniques and contribute to optimizing the design the multicore quantum computers. The presented results particularly indicate the significant role of the qubit-cavity coupling strength $g$ to ensure effective state transmission channels, while maintaining practically low $\kappa$ and $\gamma$ rates. In preliminary results, we delimit the optimal performance regions of the coupling strength $g$ and the cavity decay rate $\kappa$ that produce viable efficiency levels while minimizing operation latency and process infidelity. These design requirements are defined by a narrow operational range for the cavity decay rate $\kappa$, estimated at around $10^4$ Hz, within a system functioning in the strong coupling regime. Additionally, higher values of the coupling strength $g$ correlate directly with enhanced system performance.

According to the presented comparative analysis involving contemporary proposals for building quantum interconnects, current techniques do not reach the efficiency requirements for a reliable and extended inter-core state routing networks across the considered qubit technologies. This is mainly showcased in the studied systems, which predominantly fall within the intermediate and near-strong coupling regimes, while our analysis emphasizes on operating within the strong coupling regime for an optimal system performance. This advancement can be achieved by continuing efforts to minimize losses in qubits and/or cavities, or by focusing on enhancing the coupling strength. The sensitivity analysis presented further highlights the critical role of strong coupling in maximizing the effectiveness of quantum interconnects, which could potentially suppress the impact of lossy qubits.

Fundamentally, the coupling strength $g$ is directly dependent on the the electric dipole moment between the ground and excited states of the qubit $\mu_{g,e}$, the qubit frequency $\omega_{g,e}$, $A_{eff}$ representing the effective mode area of the cavity, and $L$ the cavity length \cite{b24}:
    \begin{equation}
        g =  \sqrt{\frac{\mu_{g,e}^2 \omega_{g,e}}{2 \epsilon_0 \hbar A_{eff} L}}
    \end{equation} \label{FM}
This suggests that a strong qubit-cavity interaction can be achieved with the following configuration: a large dipole moment $\mu_{g,e}$ and a resonant frequency between the qubit and the energy of the photons in the cavity mode. Additionally, the interaction is typically stronger in cavities where the mode volume $A_{eff}$ is smaller. Such parameters are crucial to take into account for the purpose of building fast and high-fidelity cavity-mediated networks for modular quantum computers.

For a more thorough design-oriented DSE analysis, the figure of merit may incorporate metrics of spatial constraints, and can be adapted to particularities of teleportation-based networks which efficiency is usually determined by inter-module entanglement generation rate, the fidelity of the generated entanglement, and the reconfiguration between the modules.

\begin{figure}[htbp]
\centerline{\includegraphics[scale=0.38]{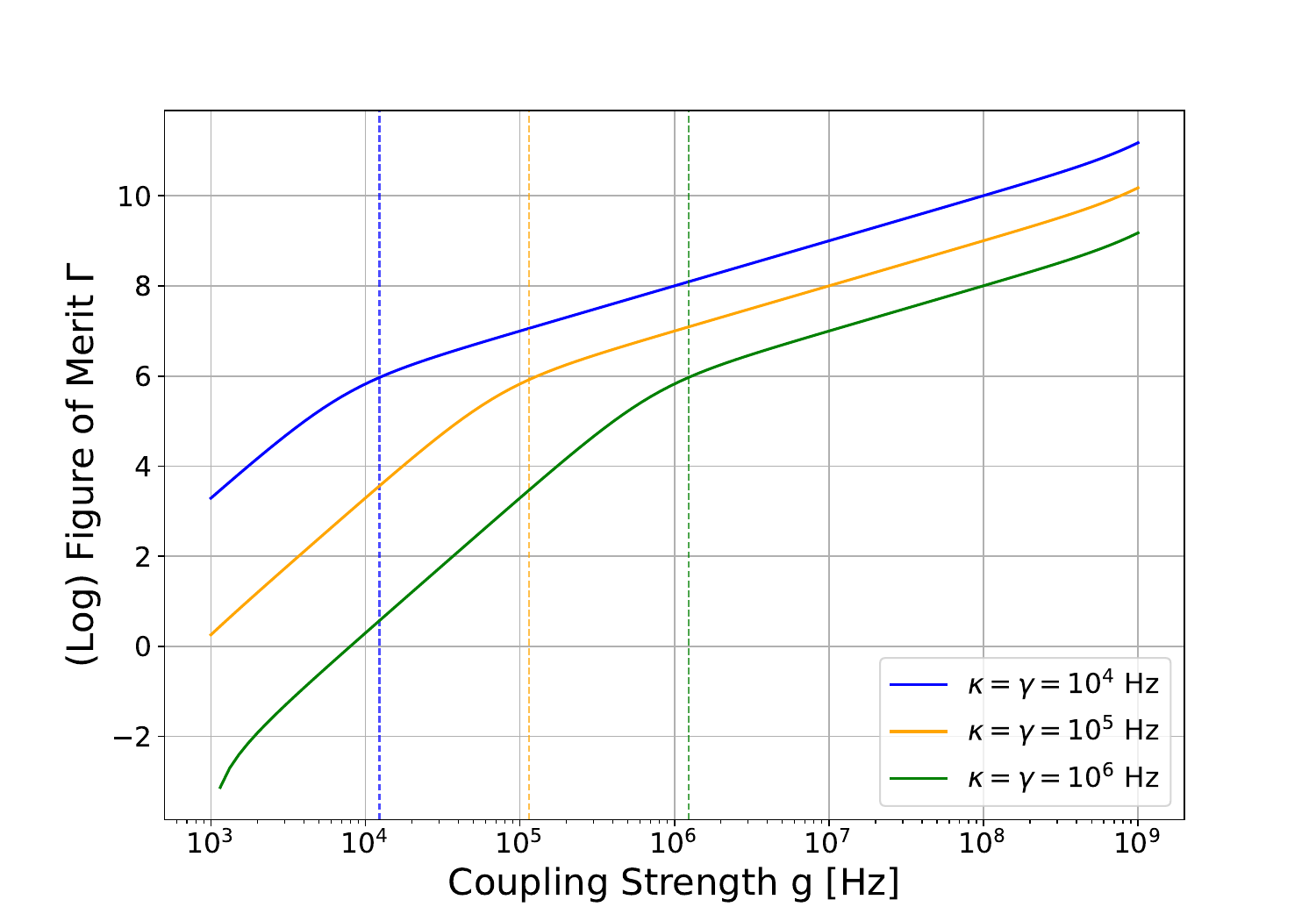}}
\caption{The variation of figure of merit $\Gamma$ with respect to coupling strength $g$ for distinct ranges of cavity decay rate $\kappa$ and atomic decay rate $\gamma$.}
\label{sa1}
\end{figure}

%% file: conclusion.tex
\section{Conclusion}
This exploratory work presents an analysis of decay rates and coupling strength in quantum cavity-mediated interconnects designed for modular quantum computers. We show the significance of the qubit-cavity coupling strength $g$ in establishing efficient transfer channels while maintaining low cavity $\kappa$ and atomic $\gamma$ decay rates. We indicate that current interconnect technologies fall short of the necessary efficiency thresholds for large-scale and reliable networks. This work also benchmarks existing interconnect proposals and offers a perspective for future design improvements based on a proposed figure of merit involving the efficiency of the quantum state transfer, the latency and infidelity of the interlink, emphasizing the importance of strong coupling to counteract qubit losses and improve state transfer fidelity.